\documentclass{article} 
\usepackage[utf8]{inputenc}
\usepackage[T1]{fontenc}
\usepackage{mathptmx}
\usepackage{etoolbox}
\usepackage{natbib}
\usepackage{relsize}
\usepackage{amssymb}
\usepackage{amsmath}
\usepackage{amsthm}
\usepackage{subcaption}
\usepackage{float}
\usepackage{hyperref}
\usepackage{adjustbox}
\usepackage{xcolor}
\usepackage{soul}
\usepackage{caption}
\usepackage[a4paper, margin=1.5in]{geometry}
\usepackage{graphicx}
\usepackage[percent]{overpic}

\title{Gravity-Driven Eco-Epidemiological Dynamics in Tri-Trophic Food Chains}

\author{
Yuvrajsingh Pravinsingh Patil$^{1}$, Eckehard Schöll $^{2}$,
Fakhteh Ghanbarnejad$^{3,*}$,
Chandrakala Meena$^{1,*}$
\\[0.5em]
$^{1}$Indian Institute of Science Education and Research (IISER) Pune, India \\
$^{2}$ Institute for Physics and Astronomy, Technische Universität Berlin, Germany \\
$^{3}$School of Technology and Architecture, SRH University of Applied Sciences Heidelberg,\\ Leipzig Campus, Germany\\
$^{*}$These authors contributed equally to this work.\\
fakhteh.ghanbarnejad@gmail.com and chandrakala@iiserpune.ac.in
\\[0.5em]
}

\date{}

\begin{document}

\maketitle
\begin{abstract}
Ecological communities are shaped by the interplay between trophic interactions and infectious disease, yet how spatially mediated interactions influence disease-driven ecosystem dynamics remains poorly understood. Here, we develop a gravity-based eco-epidemiological framework for a tri-trophic food chain in which trophic interaction depends on species abundances and effective interaction distance. The disease-free food chain system supports a stable coexistence equilibrium, providing a baseline for investigating disease-induced ecological transitions. Introducing infection at the intermediate trophic level destabilizes this equilibrium through a Hopf bifurcation, leading to sustained oscillations, whereas infection at the top predator level results in a qualitatively different transition from persistence to extinction. By systematically varying the gravity coupling strength, we show that gravity-mediated trophic interactions regulate the thresholds separating these ecological regimes, while the trophic position of infection determines the nature of the transition. Together, these findings establish a unified framework for understanding how spatially mediated trophic interactions and infectious disease jointly govern ecosystem stability, providing new insights into disease-driven dynamics in ecological communities.\\

\end{abstract}

\section{Introduction}
Ecological communities are sustained by trophic interactions that regulate the transfer of energy and biomass across multiple trophic levels \cite{LINDEMAN1942,SLOBODKIN1960,HAIRSTON1960,OKSANEN1981}. The collective outcome of these interactions determines whether populations coexist, oscillate, or collapse, making the prediction of ecological persistence and stability a central challenge in theoretical ecology and complex systems science \cite{KOT2001,MURRAY2002}. Since the pioneering predator--prey models of Lotka \cite{LOTKA1925} and Volterra \cite{VOLTERRA1928}, mathematical descriptions of food chains have progressively incorporated increasingly realistic ecological processes, including nonlinear functional responses \cite{HOLLING1959,HOLLING1959b,ROSENZWEIG1963}, trophic regulation \cite{FREEDMAN1985}, omnivory \cite{KRATINA2012}, adaptive behavior \cite{KONDOH2003, MOUGI2022}, and additional trophic levels \cite{ALANHASTINGS1991,KUZNETSOV1996,MAITI2006}. In particular, Holling's functional response theory \cite{HOLLING1959,HOLLING1959b} and the Rosenzweig-MacArthur model \cite{ROSENZWEIG1963} demonstrated that predator-prey systems can exhibit oscillatory instabilities and the paradox of enrichment, while extensions to tri-trophic food chains revealed richer dynamical behaviors, including bistability \cite{DEROSSI2015,PATTANAYAK2021}, chaos \cite{ALANHASTINGS1991,KUZNETSOV1996,SAMANTA2013}, and trophic cascades \cite{CAMPILLAY2022}. Together, these developments have established tri-trophic food-chain models as a fundamental framework for understanding nonlinear ecological dynamics and ecosystem stability.

Despite the considerable development of food-chain models, trophic interactions have almost exclusively been described using classical predator-prey interaction laws, including the Lotka-Volterra model \cite{LOTKA1925,VOLTERRA1928}, Holling-type functional responses \cite{HOLLING1959b}, the Beddington-DeAngelis functional response \cite{BEDDINGTON1975,DEANGELIS1975}, and ratio-dependent formulations \cite{ARDITI1989}. While these approaches have substantially advanced our understanding of ecological dynamics, they generally describe trophic interactions through local predator-prey relationships and do not explicitly incorporate the influence of effective ecological distance, despite the fact that species interactions in natural ecosystems are strongly shaped by dispersal \cite{LEVIN1992,OKUBO2001}, habitat connectivity \cite{TILMAN1997,WITH1997}, and spatial organization through metacommunity processes \cite{LEIBOLD2004,GUICHARD2017,ANDERSON2021}, which collectively regulate encounter rates and biomass transfer. Gravity-based interaction laws provide a natural framework for incorporating these effects by allowing interaction strength to depend jointly on population abundance and effective ecological distance. Originally developed to describe spatial interactions in human geography \cite{ZIPF1949}, transportation systems \cite{QUANDT1966}, and economics through the theoretical foundation of the gravity equation \cite{ANDERSON1979}, gravity models were subsequently generalized into a comprehensive framework for spatial interaction analysis \cite{HAYNES1985}. They have since been applied to diverse ecological processes, including biological invasions \cite{JAREMO2009}, dispersal dynamics \cite{JONGEJANS2015}, colonization probabilities of non-indigenous species \cite{HERBORG2007}, and distance-decay relationships in biodiversity \cite{MORLON2008}. More recently, Banerjee  \emph{et al.}~\cite{BAN16} also studied distant-dependent power-law interactions between ecological populations in a network of Rosenzweig–MacArthur systems.  They showed that the variation of the power-law exponent of the long-range interaction mediates transitions between full spatial synchrony and various chimera patterns, i.e., partial synchronization patterns with coexisting synchronized and desynchronized spatial domains. 
 
In addition to trophic interactions, infectious disease has emerged as another major regulator of ecological communities. Since the pioneering work of Kermack and McKendrick \cite{KERMACK1927} and the eco-epidemiological framework introduced by Anderson and May \cite{ANDERSON1986}, numerous studies have demonstrated that disease can fundamentally modify food-chain dynamics. Parasitic infection and trophic transmission alter trophic interactions and predator persistence \cite{HADELER1989}, disease modifies coexistence conditions in predator--prey systems \cite{VENTURINO1994,VENTURINO2002}, induces sustained oscillations through Hopf bifurcations \cite{CHATTOPADHYAY1999}, regulates disease persistence via epidemic thresholds \cite{HETHCOTE2004}, and suppresses chaotic dynamics while promoting stable limit cycles \cite{DAS2009}. In tri-trophic food chains, epidemiological processes introduce additional feedback because infection simultaneously influences multiple trophic levels. Consequently, disease can generate richer dynamical behavior than in two-species systems. For example, De Rossi \emph{et al.} \cite{DEROSSI2015} showed that disease affecting the primary trophic level induces persistent oscillations and bistability, whereas Mougi \cite{MOUGI_2022} demonstrated that parasite-induced changes in trophic interaction strength substantially modify food-web stability. Saveh and Ghanbarnejad showed that incorporating infectious disease dynamics into a three-level food chain can induce emergent population oscillations, alter species persistence, and, under certain conditions, drive apex predator extinction (\cite{HOOMAN2026}).

Although the present study considers a minimal deterministic tri-trophic food chain, the proposed gravity-based framework is sufficiently general to be extended to more realistic ecological systems. Future work could incorporate spatial heterogeneity, diffusion based movement of species between two-habitats, environmental stochasticity, adaptive behavior, multiple pathogens \cite{CAI2015}, or complex food-web architectures to investigate how gravity-mediated trophic interactions influence ecosystem stability under more realistic ecological conditions. Such extensions would facilitate direct comparisons with empirical systems and help assess the broader applicability of the framework.

Here, we develop a gravity-based eco-epidemiological framework for a tri-trophic food chain in which trophic interactions are governed by a gravity law. We first establish the dynamics of the disease-free system and then investigate how infection introduced at different trophic levels modifies ecosystem behavior. Specifically, we examine how disease affecting the intermediate consumer and the top level predator produces distinct ecological transitions and how the gravity coupling strength influences the thresholds separating stable coexistence, sustained oscillations, and predator persistence. By combining analytical and numerical approaches, this framework provides a unified basis for investigating the interplay between gravity-mediated trophic interactions and eco-epidemiological dynamics in ecological communities.

\section{Model description}
\label{Model_description}

The eco-epidemiological system considered in this work consists of a gravity-based tri-trophic food chain coupled with compartmental epidemic dynamics. The general form of a gravity-based interaction between species $i$ and $j$ is given by \cite{ANDERSON1979} :

\begin{equation}
\mathcal{I}_{i,j}(G_{ij},x_i,x_j,\alpha_i,\alpha_j,a_{ij},\delta_{ij})
=
G_{ij}\,
\frac{x_i^{\alpha_i}x_j^{\alpha_j}}
{a_{ij}^{\delta_{ij}}},
\label{eq:gravity_interaction}
\end{equation}

where $G_{ij}$ is a gravity-based interaction constant between species $i$ and $j$, encoding species-specific properties and environmental factors such as temperature, movement, and migration. Unless otherwise stated, the gravity coupling strength is assumed to be identical for all trophic interactions, i.e., $G_{ij} = G$.
The variables $x_i$ and $x_j$ denote species abundances, $a_{ij}$ represents the effective distance between them, $\alpha_i$ and $\alpha_j$ control how interactions scale with abundance, and $\delta_{ij}$ determines how rapidly interaction strength decays with distance. 
The resulting population dynamics is described by
\begin{equation}
\begin{aligned}
\dot{x_{1}} &=
x_1\left(b_1-d_1x_1\right)
-
c_{12}\mathcal{I}_{1,2},
\\[0.5em]
\dot{x_{2}} &=
-d_{2}x_{2}
+
c_{21}\mathcal{I}_{2,1}
-
c_{23}\mathcal{I}_{2,3},
\\[0.5em]
\dot{x_{3}} &=
-d_{3}x_{3}
+
c_{32}\mathcal{I}_{3,2},
\end{aligned}
\label{eq:gravity_model}
\end{equation}

The primary resource $x_1$ exhibits logistic growth with intrinsic growth rate $b_1$ and intraspecies competition coefficient $d_1$, while being consumed by the intermediate consumer. The intermediate consumer gains biomass through predation on the primary resource and loses biomass through natural mortality and predation by the top predator. The top predator grows by consuming the intermediate consumer and experiences natural mortality. Biomass transfer between adjacent trophic levels is mediated by the gravity-based interaction terms, with $c_{ij}$ denoting the corresponding trophic conversion efficiencies.
Infection at the secondary trophic level is incorporated through the model given in Eq.~(\ref{eq:second_species_infected})(\ref{ Secondary_model}), while the corresponding formulation for infection at the tertiary trophic level is presented in Eq.~\eqref{eq:third_species_infected}(\ref{ Tertiary_model}). Throughout the study, trophic interactions are represented using the gravity-based interaction function \(\mathcal{I}_{i,j},\)(refer Eq. \eqref{eq:gravity_interaction})  which determines the interaction strength between species through their abundances and effective interaction distance.

The analytical and numerical investigations presented in this work are based on these governing equations. Unless stated otherwise, all variables, parameters, and interaction coefficients retain the definitions introduced while introducing the models. Unless otherwise stated, all parameter values used throughout this study are listed in Table \ref{tab:parameters}. Different parameter sets were employed for the infection-free model, the secondary trophic infection model, and the tertiary trophic infection model, respectively.

For the generalized gravity-based interaction terms, we adopt the canonical exponents \(\alpha_i=\alpha_j=1\) and \(\delta_{ij}=2\). This choice recovers the classical bilinear mass-action kinetics with inverse-square distance decay, paralleling both ecological mass-action models and the traditional gravity formulations in geography and social physics. The exponent value of \(\alpha=1\) corresponds to the isometric case in ecological allometric scaling theory, distinguishing linear dependence from sub- or super-linear relationships \cite{PABLO2005}. The distance-decay exponent \(\delta=2\) aligns with the inverse-square law commonly used in spatial interaction models, which is supported by empirical and theoretical grounding in gravity modeling \cite{CHEN2015}. This classical choice offers both analytical tractability and interpretability, while serving as a useful baseline against which more complex nonlinear or distance-sensitive models may be compared.

\begin{table*}[t]
\centering
\caption{Model parameter values used in Eqs.~(\ref{eq:gravity_model}), (\ref{eq:second_species_infected}), and (\ref{eq:third_species_infected}).}
\label{tab:parameters}

\renewcommand{\arraystretch}{1.2}
\setlength{\arrayrulewidth}{0.8pt}

\begin{tabular}{|c|c|c|c|c|c|c|c|}
\hline
\textbf{Parameter} & \textbf{Value} &
\textbf{Parameter} & \textbf{Value} &
\textbf{Parameter} & \textbf{Value} &
\textbf{Parameter} & \textbf{Value} \\
\hline\hline

$b_1$ & 0.12 &
$d_1$ & 0.0002 &
$d_2$ & 0.27 &
$d_3$ & 0.20 \\
\hline

$c_{12}$ & 0.002 &
$c_{21}$ & 0.0031 &
$c_{23}$ & 0.006 &
$c_{32}$ & 0.0041 \\
\hline

$d_{2S}$ & 0.27 &
$d_{2I}$ & 0.27 &
$d_{2R}$ & 0.27 &
$d_{3S}$ & 0.20 \\
\hline

$d_{3I}$ & 0.20 &
$d_{3R}$ & 0.20 &
$c_{12}(S)$ & 0.002 &
$c_{12}(I)$ & 0.002 \\
\hline

$c_{12}(R)$ & 0.002 &
$c_{21}(S)$ & 0.0031 &
$c_{21}(I)$ & 0.0031 &
$c_{21}(R)$ & 0.0031 \\
\hline

$c_{23}(S)$ & 0.006 &
$c_{23}(I)$ & 0.006 &
$c_{23}(R)$ & 0.006 &
$c_{32}(S)$ & 0.0041 \\
\hline

$c_{32}(I)$ & 0.0041 &
$c_{32}(R)$ & 0.0041 &
& &
& \\
\hline

\end{tabular}
\end{table*}

\section{Results}
\label{Results_section}

\subsection{Gravity-based tri-trophic food chain}

\begin{figure}[htbp]
    \centering
    \includegraphics[width=0.9\linewidth]{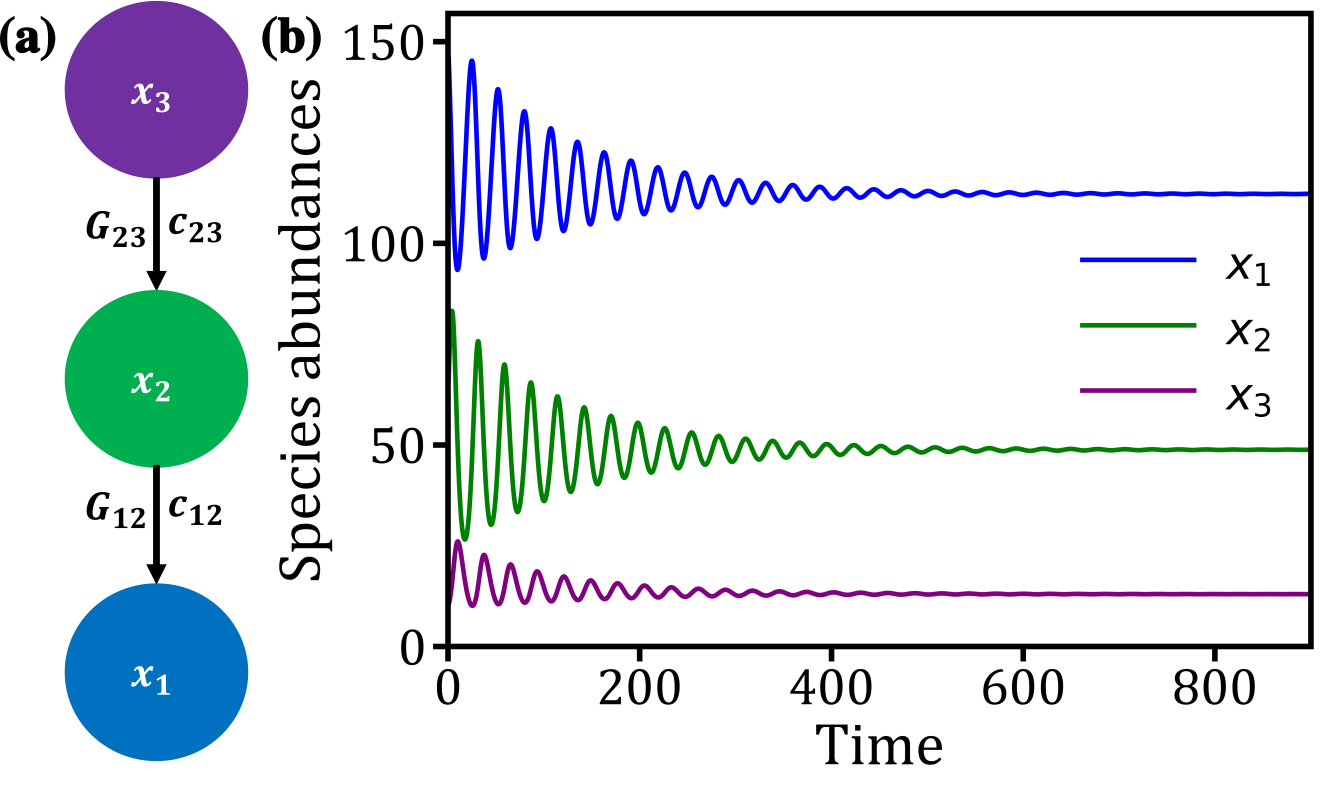}
     \caption{
    \textbf{(a)} Schematic for the tri-trophic food chain. Arrows show the direction of hunting and the parameters affecting the hunt ($c_{12}$, $c_{23}$, $G_{12}$, and $G_{23}$).
    \textbf{(b)} Time series of the species abundance for finite initial conditions . Other parameters have been explicitly mentioned in Table~\ref{tab:parameters}.}

    \label{fig:infection_less}
\end{figure}

To investigate how gravity-mediated trophic interactions influence food-chain dynamics, we first consider a gravity-based tri-trophic food chain comprising a primary resource ($x_1$), an intermediate consumer ($x_2$), and a top predator ($x_3$). Interactions between adjacent trophic levels ($i$ and $j)$ are governed by a gravity-based interaction law whose interaction term ($\mathcal{I}_{ij}$) depends jointly on population abundance and effective ecological distance (~\ref{Model_description}) \cite{ANDERSON1979}. The corresponding tri-trophic interaction is linear with the intermediate consumer feeding on the primary resource and the top predator feeding on the intermediate consumer (Fig.~\ref{fig:infection_less}(a)). Numerical analysis show that all populations converge to a stable steady states (Fig.~\ref{fig:infection_less}(b)). This observation is supported by the analytical result and also show that the steady state is both locally and globally asymptotically stable (\ref{Global_stability_proof}), providing the infection-free reference dynamical state for the subsequent eco-epidemiological analysis.

\subsubsection{Global stability proof}
\label{Global_stability_proof}

We study the reduced, infection-free system mentioned in Eq. \eqref{eq:gravity_model}. For notational compactness we dropped the explicit $(S)$ label on the $c_{ij}$ parameters; the results below hold for the susceptible-only reduced model. All parameters are assumed nonnegative with $a_{ij}>0$, $G_{ij}>0$, $d_i>0$, $b_1>0$. Below we give standard dynamical properties (positivity, boundedness, existence of equilibria, local stability) and then a global stability result under a physically interpretable reciprocity assumption.

\paragraph{Positivity and forward invariance}

If the initial state satisfies
\[
x_1(0),x_2(0),x_3(0)\ge0,
\]
then the solution remains nonnegative for all forward time:
\[
x_i(t)\ge0,\qquad t\ge0.
\]

Each equation is of the form
\[
\dot x_i=f_i(x),
\]
where \(f_i\) is continuous and satisfies
\[
f_i(x)\ge-Kx_i
\]
for some \(K\) when \(x_i\) is near \(0\) (inspection of Eq.~(\ref{eq:gravity_model}). Standard results (or the fact that the right-hand side is Lipschitz on \(\mathbb{R}_+^3\)) imply that solutions starting on the nonnegative orthant cannot cross into negative values: if some coordinate tends to \(0\), its derivative at zero is nonnegative (because all terms that can be negative are proportional to that coordinate), so it cannot decrease through zero. Thus \(\mathbb{R}_+^3\) is forward invariant.

\paragraph{Boundedness and absorbing set}

All solutions with nonnegative initial data are bounded and enter a compact absorbing set.
From Eq.~(\ref{eq:gravity_model}),
\[
\dot x_1 \le x_1(b_1-d_1x_1),
\]
so \(x_1(t)\) is bounded above by the logistic solution and thus satisfies
\[
x_1(t)\le
\max\left\{
x_1(0),\frac{b_1}{d_1}
\right\}.
\]

For \(x_2\) and \(x_3\), observe

\[
\dot x_2
\le
-d_2x_2
+
c_{21}G_{12}
\frac{x_1x_2}{a_{12}^2}
\le
\left(
-d_2+
c_{21}G_{12}
\frac{\sup x_1}{a_{12}^2}
\right)x_2,
\]

so either the parenthesis is negative, which yields exponential decay (hence boundedness), or it is positive but then growth of \(x_2\) is limited by predation and eventual feedback to \(x_1\) (which itself is bounded). A similar estimate holds for \(x_3\). A standard comparison argument gives a global bound and a compact absorbing set in \(\mathbb{R}_+^3\); thus all trajectories enter and remain in some compact region.

\paragraph{Existence of equilibria}
The equilibrium points of the reduced infection-free system are obtained by setting
\[
\dot{x}_1=\dot{x}_2=\dot{x}_3=0.
\]
The system admits the following equilibrium solutions.

\paragraph{Extinction equilibrium.}
The trivial equilibrium is
\[
E_0=(0,0,0),
\]
corresponding to extinction of all three species.

\paragraph{Primary-species equilibrium.}

If the intermediate and tertiary species are absent, the primary species approaches its carrying capacity, giving
\[
E_1=\left(\frac{b_1}{d_1},\,0,\,0\right).
\]

\paragraph{Coexistence equilibrium.}
For the interior equilibrium,
\[
E^*=(x_1^*,x_2^*,x_3^*),
\]

all three populations are positive and satisfy

\begin{subequations}
\label{eq:A_steady_state}
\begin{align}
0&=
x_1^*
\left(b_1-d_1x_1^*\right)
-
c_{12}G_{12}
\frac{x_1^*x_2^*}{a_{12}^{2}},
\\
0&=
-d_2x_2^*
+
c_{21}G_{12}
\frac{x_1^*x_2^*}{a_{12}^{2}}
-
c_{23}G_{23}
\frac{x_2^*x_3^*}{a_{23}^{2}},
\\
0&=
-d_3x_3^*
+
c_{32}G_{23}
\frac{x_2^*x_3^*}{a_{23}^{2}}.
\end{align}
\end{subequations}

Assuming

\[
x_1^*>0,\qquad
x_2^*>0,\qquad
x_3^*>0,
\]

the equilibrium coordinates satisfy

\[
x_2^*
=
\frac{d_3a_{23}^{2}}
{c_{32}G_{23}},
\]

\[
x_1^*
=
\frac{
b_1
-
\dfrac{c_{12}G_{12}}{a_{12}^{2}}x_2^*
}
{d_1},
\]

and

\[
x_3^*
=
\frac{a_{23}^{2}}
{c_{23}G_{23}}
\left(
\frac{c_{21}G_{12}}{a_{12}^{2}}x_1^*
-
d_2
\right).
\]

Hence, an interior equilibrium exists whenever all three equilibrium populations are positive.

\paragraph{Local stability of the coexistence equilibrium}
To determine the local stability of the interior equilibrium
\(
E^*=(x_1^*,x_2^*,x_3^*)
\),
we linearize the dynamical system about \(E^*\). The corresponding Jacobian matrix is

\begin{equation}
J(E^*)=
\begin{pmatrix}
\displaystyle
b_1-2d_1x_1^*
-\frac{c_{12}G_{12}}{a_{12}^{2}}x_2^*
&
\displaystyle
-\frac{c_{12}G_{12}}{a_{12}^{2}}x_1^*
&
0
\\[1.2em]

\displaystyle
\frac{c_{21}G_{12}}{a_{12}^{2}}x_2^*
&
\displaystyle
-d_2
+\frac{c_{21}G_{12}}{a_{12}^{2}}x_1^*
-\frac{c_{23}G_{23}}{a_{23}^{2}}x_3^*
&
\displaystyle
-\frac{c_{23}G_{23}}{a_{23}^{2}}x_2^*
\\[1.2em]

0
&
\displaystyle
\frac{c_{32}G_{23}}{a_{23}^{2}}x_3^*
&
\displaystyle
-d_3
+\frac{c_{32}G_{23}}{a_{23}^{2}}x_2^*
\end{pmatrix}.
\end{equation}

The characteristic polynomial associated with the Jacobian is

\[
\lambda^3+a_1\lambda^2+a_2\lambda+a_3=0,
\]

where the coefficients \(a_1\), \(a_2\), and \(a_3\) are determined by the trace, principal minors, and determinant of the Jacobian matrix.

According to the Routh--Hurwitz criterion, the coexistence equilibrium is locally asymptotically stable if
\[
a_1>0,\qquad
a_2>0,\qquad
a_3>0,
\]
and
\[
a_1a_2>a_3.
\]

These conditions ensure that all eigenvalues of the Jacobian possess negative real parts, thereby guaranteeing local asymptotic stability of the coexistence equilibrium.

\paragraph{Global asymptotic stability}
To establish global convergence of the coexistence equilibrium, we construct a Lyapunov function for the reduced infection-free system under a reciprocity (detailed balance) condition relating the trophic interaction coefficients. This condition ensures that the biomass transfer between adjacent trophic levels satisfies a symmetric energy balance, allowing the nonlinear interaction terms to cancel in the Lyapunov derivative.

Consider the Lyapunov function
\begin{equation}
V(x_1,x_2,x_3)
=
\left(
x_1-x_1^*
-
x_1^*\ln\frac{x_1}{x_1^*}
\right)
+
A
\left(
x_2-x_2^*
-
x_2^*\ln\frac{x_2}{x_2^*}
\right)
+
B
\left(
x_3-x_3^*
-
x_3^*\ln\frac{x_3}{x_3^*}
\right),
\end{equation}

where \(A>0\) and \(B>0\) are constants chosen so that the mixed interaction terms cancel in the time derivative of \(V\).

Differentiating \(V\) along the trajectories of the system gives
\begin{equation}
    \dot V
=
\sum_{i=1}^{3}
\left(
1-\frac{x_i^*}{x_i}
\right)
\dot x_i.
\end{equation}
Using the equilibrium relations (\ref{eq:A_steady_state}) , the interaction terms cancel identically, leaving only quadratic terms associated with the logistic self-regulation of the primary species. Consequently,
\begin{equation}
 \dot V\le0,   
\end{equation}
with equality holding only at the coexistence equilibrium. Therefore, by LaSalle's invariance principle, every trajectory beginning in the positive invariant region converges to the unique interior equilibrium. Hence $\dot V\le 0$ for all $x\in\mathbb{R}_+^3$, and $\dot V=0$ if and only if $x=x^*$. By LaSalle’s invariance principle the interior equilibrium is globally asymptotically stable in the positive orthant.
\qed

\subsection{Infection at the intermediate trophic level}

To investigate how infectious disease modifies gravity-mediated trophic interactions and dynamics of the species, we extend the infection-free food chain by introducing infection into the intermediate consumer while keeping the primary resource and the top predator disease free (refer \ref{ Secondary_model} and Eq.~\eqref{eq:second_species_infected}). The intermediate population is divided into susceptible ($x_{2S}$), infected ($x_{2I}$), and recovered ($x_{2R}$) compartments (Fig.~\ref{fig:oscillation_panels}(a)). Infection modifies trophic interactions through the infection-modulation parameter $w$, reducing the feeding efficiency of infected individuals on the primary resource while increasing their vulnerability to predation by the top predator, thereby redistributing biomass transfer across trophic levels.

\subsubsection{Eco-epidemiological model with infection at the intermediate trophic level}
\label{ Secondary_model}

\begin{figure}[htbp]
    \centering
    \includegraphics[width=0.9\linewidth]{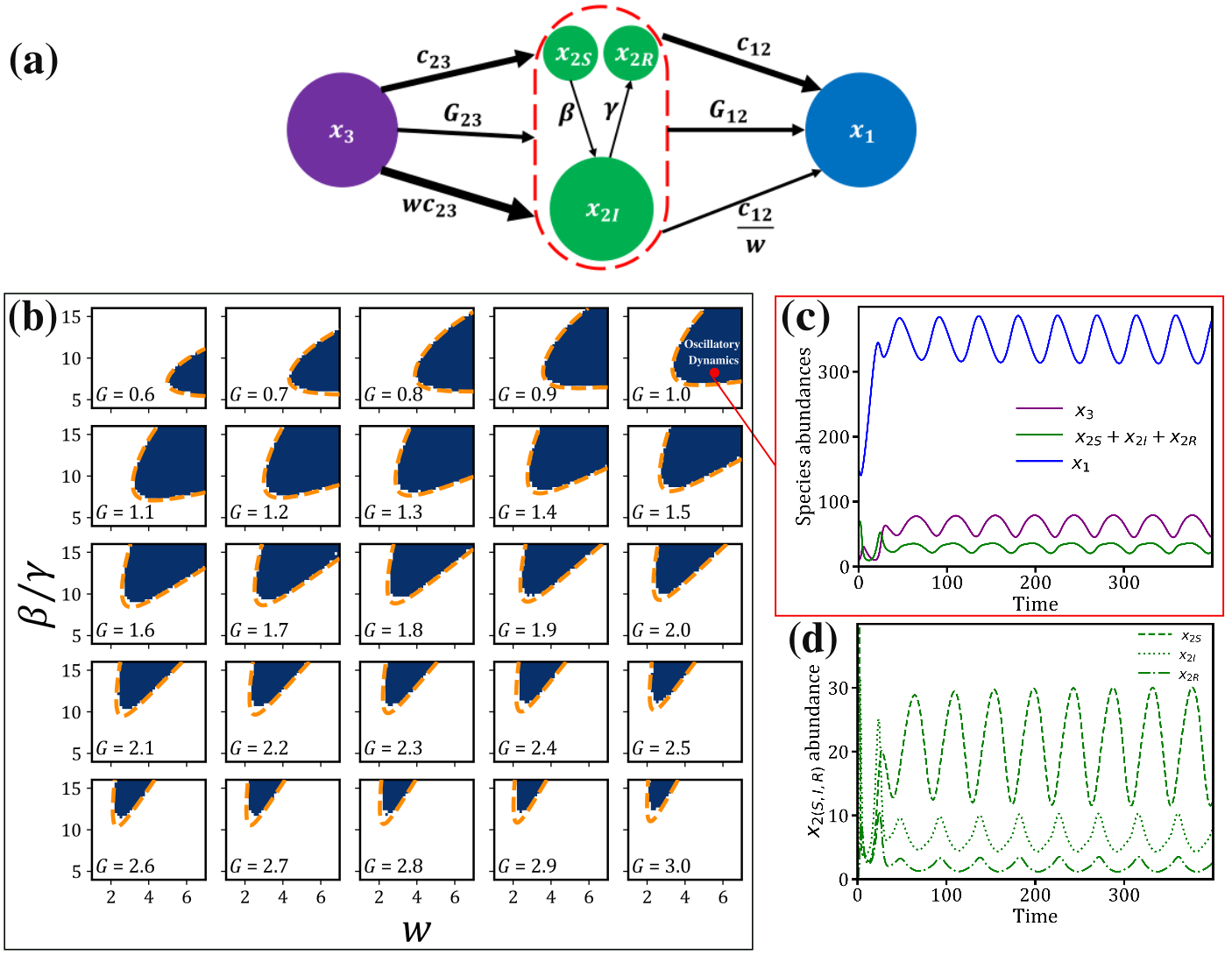}
   \caption{
\textbf{(a)}Schematic representation of the three species food-chain model with infection introduced at the secondary trophic level\textbf{ $x_2$.} The secondary population is divided into susceptible ($x_{2S}$), infected ($x_{2I}$), and recovered ($x_{2I}$) classes, which interact with the primary ($x_1$) and the tertiary $x_3$ species underlying gravity-based trophic interactions. The infection-modulated predation parameter $w$ \cite{HOOMAN2026} alters these interactions by reducing the feeding efficiency of infected individuals on the primary species while increasing their susceptibility to predation by the tertiary species.\textbf{(b)} Dynamical regime diagram in the $\beta/\gamma$ and $w$ parameter space, where blue regions denote sustained oscillatory dynamics, whereas white regions correspond to convergence to the stable coexistence equilibrium. As $G$ increases, the oscillatory regime systematically shifts toward lower values of $w$ and higher values of $\beta/\gamma$, indicating that gravity-mediated interactions systematically modify the parameter range over which sustained oscillations occur.
\textbf{(c)} Representative time series of the primary resource ($x_1$), the total intermediate population ($x_{2}^{\mathrm{total}} = x_{2S} + x_{2I} + x_{2R}$), and the top level species ($x_3$) for a representative parameter set within the oscillatory regime ($G=1$, $\beta=10$, $\gamma=0.2$, and $w=5$). All three trophic levels exhibit sustained periodic oscillations.
\textbf{(d)} Corresponding temporal evolution of the susceptible ($x_{2S}$), infected ($x_{2I}$), and recovered ($x_{2R}$) compartments of the intermediate populations for the same parameter values, illustrating persistent oscillations in the epidemiological dynamics.
\textbf{(e)} Analytical prediction of the oscillatory regime in the $(w,\beta/\gamma)$ parameter space for $G=1$, obtained from the analytical conditions given by Eq.~(\ref{eq:phi}. The analytical prediction shows excellent agreement with the numerical oscillatory regime in panel (b), with only small discrepancies near the boundary.
}
\label{fig:oscillation_panels}
\end{figure}

To investigate how infectious disease modifies gravity-mediated trophic interactions, we extend the infection-free tri-trophic food-chain model by introducing disease transmission within the intermediate consumer population while keeping the primary resource ($x_1$) and the top predator ($x_3$) disease free. The intermediate trophic level is partitioned into three epidemiological compartments: susceptible ($x_{2S}$), infected ($x_{2I}$), and recovered ($x_{2R}$), as illustrated in Fig.~\ref{fig:oscillation_panels}(a).

The ecological effects of infection are incorporated through an infection-modulated interaction parameter, $w$, which captures the reduced ecological fitness of infected individuals. As infection progresses, infected individuals become less efficient at consuming the primary resource while simultaneously becoming more vulnerable to predation by the top predator. Accordingly, the effective trophic interaction coefficients are modified as

\begin{equation}
c_{21}^{\mathrm{eff}}(I)=\frac{c_{21}(I)}{w},
\qquad
c_{12}^{\mathrm{eff}}(I)=\frac{c_{12}(I)}{w},
\end{equation}

for interactions between infected individuals and the primary resource, and

\begin{equation}
c_{23}^{\mathrm{eff}}(I)=wc_{23}(I),
\qquad
c_{32}^{\mathrm{eff}}(I)=wc_{32}(I),
\end{equation}

for interactions between infected individuals and the top predator. Consequently, increasing $w$ simultaneously weakens the role of infected individuals as consumers while strengthening their role as prey.

Using the gravity-based interaction defined in Eq.~(\ref{eq:gravity_interaction}), the eco-epidemiological dynamics are governed by

\begin{equation}
\begin{aligned}
\dot{x}_{1} &= x_1 \left( b_1 - d_1 x_1 \right)
- c_{12}(S) \mathcal{I}_{1,2S}
- c_{12}^{\mathrm{eff}}(I) \mathcal{I}_{1,2I}
- c_{12}(R) \mathcal{I}_{1,2R}
\\[1em]
\dot{x}_{2S} &= -d_{2S} x_{2S}
+ c_{21}(S) \mathcal{I}_{2S,1}
- c_{23}(S) \mathcal{I}_{2S,3}
-\beta\frac{x_{2S}x_{2I}}{N_2}
+c_{21}(R)\mathcal{I}_{2R,1}
\\[1em]
\dot{x}_{2I} &= -d_{2I}x_{2I}
+c_{21}^{\mathrm{eff}}(I)\mathcal{I}_{2I,1}
-c_{23}^{\mathrm{eff}}(I)\mathcal{I}_{2I,3}
-\gamma x_{2I}
+\beta\frac{x_{2S}x_{2I}}{N_2}
\\[1em]
\dot{x}_{2R} &= -d_{2R}x_{2R}
-c_{23}(R)\mathcal{I}_{2R,3}
+\gamma x_{2I}
\\[1em]
\dot{x}_{3} &= -d_3x_3
+c_{32}(S)\mathcal{I}_{3,2S}
+c_{32}^{\mathrm{eff}}(I)\mathcal{I}_{3,2I}
+c_{32}(R)\mathcal{I}_{3,2R},
\end{aligned}
\label{eq:second_species_infected}
\end{equation}

where

\[
N_2=x_{2S}+x_{2I}+x_{2R}
\]

denotes the total intermediate population.

Disease transmission follows frequency-dependent incidence, with susceptible individuals becoming infected at rate $\beta x_{2S}x_{2I}/N_2$. Infected individuals recover at rate $\gamma x_{2I}$ and enter the recovered compartment, while mortality of infected individuals is represented through the parameter $d_{2I}$. Consequently, epidemiological transitions are fully coupled to the gravity-mediated trophic interactions through the modified interaction coefficients.

\subsubsection{Infection induces sustained oscillations}
We first examine the consequences of introducing infection at the intermediate trophic level. In contrast to the infection-free system, the coexistence equilibrium loses stability, and the coupled eco-epidemiological dynamics evolve toward sustained oscillatory behavior. Following an initial transient, populations at all trophic levels exhibit persistent oscillations rather than converging to a steady state (Fig.~\ref{fig:oscillation_panels}(c)). The susceptible ($x_{2S}$), infected ($x_{2I}$), and recovered ($x_{2R}$) compartments of the intermediate consumer also oscillate persistently (Fig.~\ref{fig:oscillation_panels}(d)), demonstrating that epidemiological cycling is coupled to trophic biomass transfer and drives oscillatory behavior across all trophic levels.

To determine how epidemiological processes and gravity-mediated interactions jointly regulate the onset of oscillatory dynamics, we construct numerical dynamical regime diagrams in the $(\beta/\gamma,w)$ parameter space for different values of the gravity coupling strength $G$, where $\beta$ and $\gamma$ denote the transmission and recovery rates, respectively (Fig.~\ref{fig:oscillation_panels}(b)). Increasing $G$ systematically shifts the oscillatory regime toward lower values of the infection-modulation parameter $w$ and the higher value of the transmission-to-recovery ratio, indicating  that gravity-mediated interactions modify the combinations of transmission-to-recovery ratio and infection-modulation parameter required for the emergence of oscillatory dynamics.

The emergence of sustained oscillatory behavior indicates that the coexistence equilibrium loses stability through a Hopf bifurcation. To establish the origin of this transition, we perform a local stability analysis of the coexistence equilibrium (~\ref{Stability_analysis_complete_coefficient_criterion}). Applying the coefficient criterion of Ref.~\cite{DOUSKOS2015} yields the analytical Hopf bifurcation boundary,

\begin{equation}
\Phi=(b_3-b_1b_2)(b_5b_2-b_3b_4)-(b_5-b_1b_4)^2=0,
\label{eq:phi}
\end{equation}

where the coefficients $b_i$ are functions of the system parameters (~\ref{ Stability_analysis}). These conditions define the analytical boundary separating the stable coexistence and oscillatory regimes in the $(\beta/\gamma,w)$ parameter space. As shown in Fig.~\ref{fig:oscillation_panels}(e), the analytical boundary closely matches the numerical dynamical regime obtained for $G=1$ (Fig.~\ref{fig:oscillation_panels}(b)), demonstrating that the onset of oscillatory behavior is accurately captured by the Hopf bifurcation analysis.

\subsubsection{Analytical analysis of Emerging Oscillatory dynamics}
\label{ Stability_analysis}

To determine the onset of oscillatory dynamics, we analyze the local stability of the coexistence equilibrium of Eq.~(\ref{eq:second_species_infected}). Linearization about the equilibrium yields the Jacobian matrix, whose explicit form is provided in \ref{Stability_analysis_complete_coefficient_criterion}.

The corresponding characteristic polynomial is

\begin{equation}
\lambda^5+b_1\lambda^4+b_2\lambda^3+b_3\lambda^2+b_4\lambda+b_5=0,
\label{eq:characteristic_polynomial}
\end{equation}

where the coefficients $b_i$ are functions of the ecological and epidemiological parameters.

The stability of the coexistence equilibrium is examined using the complete coefficient criterion for five-dimensional dynamical systems \cite{DOUSKOS2015}. A Hopf bifurcation occurs when the condition \eqref{eq:phi} holds true. These conditions define the analytical boundary separating the stable coexistence and oscillatory regimes in the $(\beta/\gamma,w)$ parameter space.

\subsubsection{Stability Analysis of Eco-epidemiological model with infection at the secondary trophic level}
\label{Stability_analysis_complete_coefficient_criterion} 
The onset of oscillatory dynamics in the secondary trophic infection model was investigated by linearizing the governing equations (\eqref{eq:second_species_infected}) about the corresponding coexistence equilibrium. Denoting the state vector by
\[
X=(x_1,x_{2S},x_{2I},x_{2R},x_3),
\]
the linearized dynamics are governed by the Jacobian matrix \(J(X)\), evaluated at the equilibrium point. The Jacobian is evaluated as:

\small
\begin{equation}
J(X)=
\begin{pmatrix}

J_{11} &
-\dfrac{c_{12}(S)G_{12}}{a_{12}^{2}}x_1 &
-\dfrac{c_{12}^{\mathrm{eff}}(I)G_{12}}{a_{12}^{2}}x_1 &
-\dfrac{c_{12}(R)G_{12}}{a_{12}^{2}}x_1 &
0

\\[1.2em]

\dfrac{c_{21}(S)G_{12}}{a_{12}^{2}}x_{2S} &
J_{22} &
-\beta\dfrac{x_{2S}(x_{2S}+x_{2R})}{N_2^{2}} &
\dfrac{c_{21}(R)G_{12}}{a_{12}^{2}}x_{2R}
+\beta\dfrac{x_{2S}x_{2I}}{N_2^{2}} &
-\dfrac{c_{23}(S)G_{23}}{a_{23}^{2}}x_{2S}

\\[1.2em]

\dfrac{c_{21}^{\mathrm{eff}}(I)G_{12}}{a_{12}^{2}}x_{2I} &
\beta\dfrac{x_{2I}(x_{2I}+x_{2R})}{N_2^{2}} &
J_{33} &
-\beta\dfrac{x_{2S}x_{2I}}{N_2^{2}} &
-\dfrac{c_{23}^{\mathrm{eff}}(I)G_{23}}{a_{23}^{2}}x_{2I}

\\[1.2em]

0 &
0 &
\gamma &
J_{44} &
-\dfrac{c_{23}(R)G_{23}}{a_{23}^{2}}x_{2R}

\\[1.2em]

0 &
\dfrac{c_{32}(S)G_{23}}{a_{23}^{2}}x_3 &
\dfrac{c_{32}^{\mathrm{eff}}(I)G_{23}}{a_{23}^{2}}x_3 &
\dfrac{c_{32}(R)G_{23}}{a_{23}^{2}}x_3 &
J_{55}

\end{pmatrix},
\end{equation}

\begin{align}
J_{11}
&=
b_1
-
2d_1x_1
-
\dfrac{G_{12}}{a_{12}^{2}}
\left[
c_{12}(S)x_{2S}
+
c_{12}^{\mathrm{eff}}(I)x_{2I}
+
c_{12}(R)x_{2R}
\right],
\\[1em]
J_{22}
&=
-d_{2S}
+
\dfrac{c_{21}(S)G_{12}}{a_{12}^{2}}x_1
-
\dfrac{c_{23}(S)G_{23}}{a_{23}^{2}}x_3
-
\beta
\dfrac{x_{2I}(N_2-x_{2S})}{N_2^{2}},
\\[1em]
J_{33}
&=
-d_{2I}
-\gamma
+
\dfrac{c_{21}^{\mathrm{eff}}(I)G_{12}}{a_{12}^{2}}x_1
-
\dfrac{c_{23}^{\mathrm{eff}}(I)G_{23}}{a_{23}^{2}}x_3
+
\beta
\dfrac{x_{2S}(N_2-x_{2I})}{N_2^{2}},
\\[1em]
J_{44}
&=
-d_{2R}
-
\dfrac{c_{23}(R)G_{23}}{a_{23}^{2}}x_3
+
\dfrac{c_{21}(R)G_{12}}{a_{12}^{2}}x_1,
\\[1em]
J_{55}
&=
-d_3
+
\dfrac{G_{23}}{a_{23}^{2}}
\left[
c_{32}(S)x_{2S}
+
c_{32}^{\mathrm{eff}}(I)x_{2I}
+
c_{32}(R)x_{2R}
\right].
\end{align}

The stability properties of the equilibrium are determined by the eigenvalue spectrum of \(J(X)\), obtained from the corresponding characteristic equation. For the secondary trophic infection model, the characteristic polynomial takes the form

\begin{equation}
\lambda^5+b_1\lambda^4+b_2\lambda^3+b_3\lambda^2+b_4\lambda+b_5=0,
\end{equation}

where the coefficients \(b_i\) are explicit functions of the system parameters, including the gravity-based coupling strength \(G_{ij}\), the infection rate \(\beta\), the recovery rate \(\gamma\), the trophic interaction coefficients \(c_{ij}\), and the infection-modulated predation parameter \(w\).

The coefficients \(b_i\) are determined by the principal minors of the Jacobian matrix. Explicitly,

\begin{align}
b_{1}
&=
-\sum_{i=1}^{5}J_{ii}
\nonumber\\
&=
-\left(
J_{11}+J_{22}+J_{33}+J_{44}+J_{55}
\right),
\\[1em]
b_{2}
&=
\sum_{1\le i<j\le5}
\begin{vmatrix}
J_{ii} & J_{ij}\\
J_{ji} & J_{jj}
\end{vmatrix}
\nonumber\\
&=
\sum_{1\le i<j\le5}
\left(
J_{ii}J_{jj}
-
J_{ij}J_{ji}
\right),
\\[1em]
b_{3}
&=
-
\sum_{1\le i<j<k\le5}
\det
\begin{pmatrix}
J_{ii} & J_{ij} & J_{ik}\\
J_{ji} & J_{jj} & J_{jk}\\
J_{ki} & J_{kj} & J_{kk}
\end{pmatrix},
\\[1em]
b_{4}
&=
\sum_{1\le i<j<k<l\le5}
\det(M_{ijkl}),
\\[1em]
b_{5}
&=
-\det(J),
\end{align}

where \(M_{ijkl}\) denotes the \(4\times4\) principal submatrix obtained by retaining rows and columns \(i,j,k,\) and \(l\).
The resulting characteristic polynomial forms the basis for the complete coefficient criterion used to identify the onset of oscillatory dynamics.

\subsubsection{Gravity shifts the oscillatory boundary}

\begin{figure}[htbp]
\centering
\includegraphics[width=0.9\linewidth]{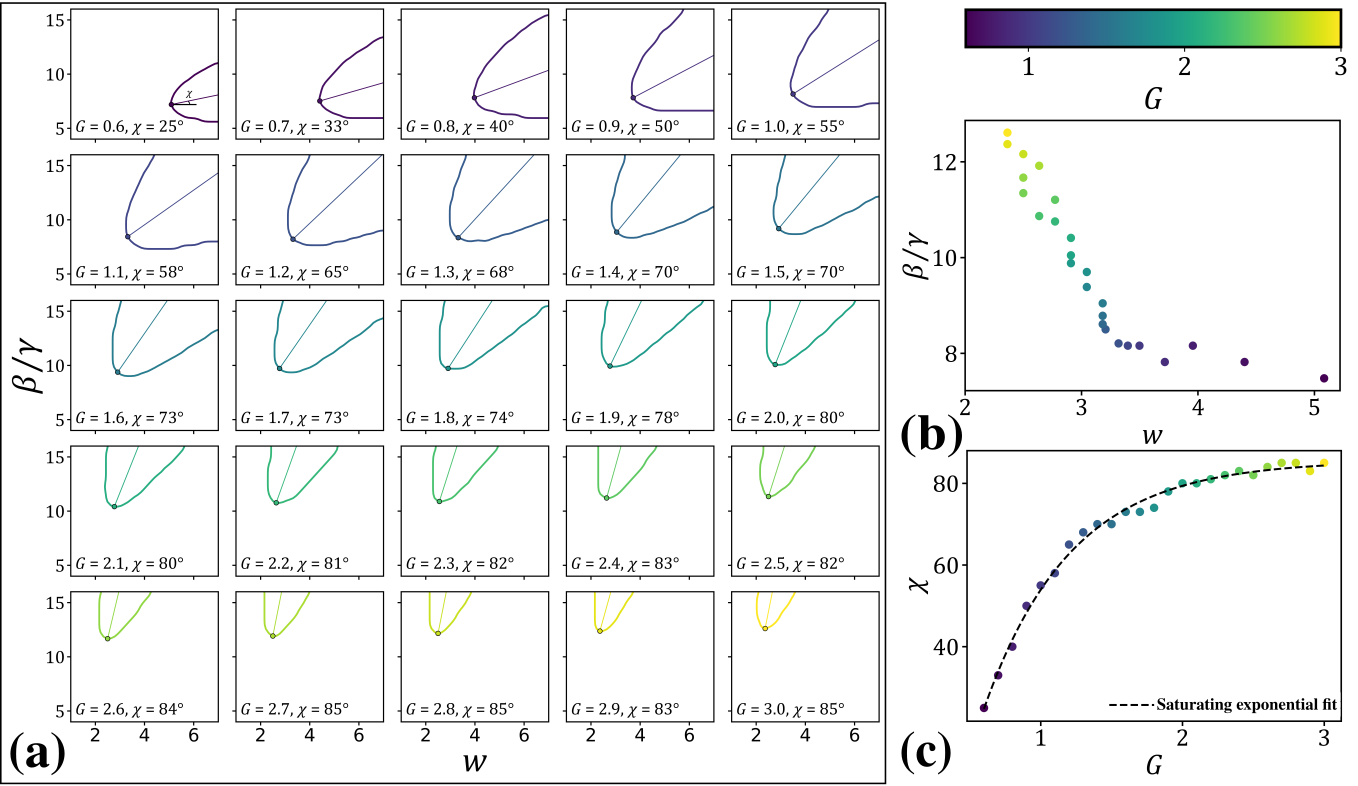}
\caption{
\textbf{(a)} Oscillatory boundaries corresponding to different values of $G$ plotted in the $(w,\beta/\gamma)$ plane. For each boundary, the knee point is identified using the maximum--distance method and marked on the curve (refer \ref{second_species_oscillatory_boundary}). A straight line passing through the knee point represents the axis of the parabolic approximation of the boundary. The angle $\chi$ between this axis and the $w$--axis is illustrated in the first subplot. The colors of the boundaries, knee points, and corresponding axes indicate the values of $G$ as shown by the color bar. 
\textbf{(b)} Color bar showing range of $G$ used to highlight the value of $G$ used in sub-panels of panel (a), panel (c) and panel (d).
\textbf{(c)} Translation of the knee points in the $(w,\beta/\gamma)$ parameter space as the parameter $G$ changes, demonstrating the systematic shift of the oscillatory boundary.  
\textbf{(d)} Rotation of the oscillatory boundaries quantified through the orientation angle $\chi$. The dependence of $\chi$ on $G$ is shown together with a fitted saturating exponential function, indicating that the rotation of the boundary approaches an asymptotic orientation as $G$ increases.
}
\label{fig:osc_boundary_geom}
\end{figure}

To quantify how gravity-mediated interactions reshape the oscillatory regime, we examine the geometry of its boundary in the $(w,\beta/\gamma)$ parameter space (Fig.~\ref{fig:osc_boundary_geom}(a)). As the gravity coupling strength increases, the oscillatory boundary undergoes a systematic translation toward lower values of $w$, indicating that oscillatory dynamics become accessible over a broader parameter range (Fig.~\ref{fig:osc_boundary_geom}(c)). In addition to this translation, the boundary progressively rotates, reflecting a continuous reorganization of the combinations of infection and trophic interaction strengths that support sustained oscillations.

To characterize this geometric change, we quantify the orientation angle $\chi$ of the oscillatory boundary ( \ref{second_species_oscillatory_boundary}). The angle increases monotonically with the gravity coupling strength and approaches an asymptotic value that is accurately described by

\begin{equation}
\chi(G)=\chi_{\infty}-Ae^{-kG},
\label{chi(G)}
\end{equation}

(Fig.~\ref{fig:osc_boundary_geom}(d)). This saturating behavior indicates that the influence of gravity on the orientation of the oscillatory boundary becomes progressively weaker at large coupling strengths, revealing a robust geometric organization of the oscillatory regime.

\subsubsection{Geometric characterization of the oscillatory boundary}
\label{ Geometric_analysis}

To quantify how gravity-mediated interactions reshape the oscillatory regime, the oscillatory boundary is extracted numerically in the $(w,\beta/\gamma)$ parameter space for different values of the gravity coupling strength $G$. The knee point of each boundary is identified using the maximum-distance method, as described in ~\ref{second_species_oscillatory_boundary}. Around the knee point, the oscillatory boundary is locally approximated by a parabola, allowing its symmetry axis to be determined. The orientation of this axis is quantified by the angle $\chi$ measured with respect to the $w$-axis.

The dependence of the orientation angle on gravity coupling is fitted using

\begin{equation}
\chi(G)=\chi_{\infty}-Ae^{-kG},
\end{equation}

where $\chi_{\infty}$ denotes the asymptotic orientation angle, and $A$ and $k$ are fitting parameters obtained by nonlinear least-squares regression.

\subsubsection{Geometry of the oscillatory boundary}
\label{second_species_oscillatory_boundary}
To quantify how the oscillatory regime changes with the gravity-based coupling strength \(G\), we performed a geometric analysis of the oscillatory boundaries obtained from the parameter-space diagrams. The objective is to characterize the systematic translation and rotation of the oscillatory boundary as the coupling strength varies. For each value of the gravity-based coupling strength \(G\), the oscillatory boundary is extracted numerically from the corresponding parameter-space map. To obtain a representative point on each oscillatory boundary, the knee point is determined geometrically. The knee point is defined as the location of maximum curvature and provides a compact description of the overall position of the oscillatory regime. This representative point is subsequently used to quantify the translation of the oscillatory boundary as the gravity-based coupling strength changes.
The oscillatory boundary is obtained numerically as a parametric curve separating oscillatory and non-oscillatory regimes in the $(w, \beta/\gamma)$ parameter plane. This boundary is represented by a set of discrete points
\[
(x_i , y_i), \quad i = 1,2,\ldots,N,
\]
where $x_i = w$ and $y_i = \beta/\gamma$.

To determine the knee point, we use the maximum distance method. Let the endpoints of the curve be
\[
P_1 = (x_1, y_1), \qquad P_2 = (x_N, y_N).
\]

The direction vector of the line connecting these points is
\[
\mathbf{v} = P_2 - P_1,
\]
and the corresponding unit vector is
\[
\hat{\mathbf{v}} = \frac{\mathbf{v}}{\|\mathbf{v}\|}.
\]

For any point $P_i = (x_i, y_i)$ on the curve, we define the displacement vector
\[
\mathbf{u}_i = P_i - P_1.
\]

The projection of $\mathbf{u}_i$ onto the line direction is given by
\[
\mathbf{p}_i = (\mathbf{u}_i \cdot \hat{\mathbf{v}})\hat{\mathbf{v}},
\]
and the perpendicular component is
\[
\mathbf{d}_i = \mathbf{u}_i - \mathbf{p}_i.
\]

The perpendicular distance from the curve to the reference line is
\[
D_i = \|\mathbf{d}_i\|.
\]

The knee point is then identified as
\[
i^{*} = \arg\max_{i} D_i,
\]
with coordinates
\[
(x_{\text{knee}}, y_{\text{knee}}) = (x_{i^{*}}, y_{i^{*}}).
\]

This procedure is repeated for each value of $G$, allowing the evolution of the oscillatory boundary to be tracked through the movement of the knee points.

The oscillatory boundary is locally approximated by a parabolic curve, and an axis passing through the knee point is constructed to characterize its orientation. The angle $\chi$ between this axis and the $w$-axis is computed for each value of $G$.

The variation of $\chi$ with $G$ is described by a saturating exponential function,
\begin{equation}
\chi(G) = \chi_{\infty} - A e^{-kG},
\end{equation}
where $\chi_{\infty}$ is the asymptotic angle, $A$ is the amplitude, and $k$ is the rate constant.

\subsection{Infection at the tertiary trophic level}
\label{tertiary_trophic_infection}
To examine the impact of infectious disease at the top of the food chain, we extend the gravity-based food-chain model by introducing infection into the tertiary trophic level while keeping the primary resource and the intermediate consumer disease free ( \ref{ Tertiary_model}). The tertiary population is divided into susceptible ($x_{3S}$), infected ($x_{3I}$), and recovered ($x_{3R}$) compartments (Fig.~\ref{fig:Third_species_schematic}(a)). Infection modifies trophic interactions through the infection-modulated interaction parameter $w$, which reduces the predation efficiency of infected apex predators. Consequently, increasing $w$ weakens trophic energy transfer from the intermediate consumer to the apex predator, thereby altering the long-term persistence of the tertiary trophic level.

\subsubsection{Eco-epidemiological model with infection at the tertiary trophic level}
\label{ Tertiary_model}

\begin{figure}[htbp]
    \centering
    \includegraphics[width=0.9\linewidth]{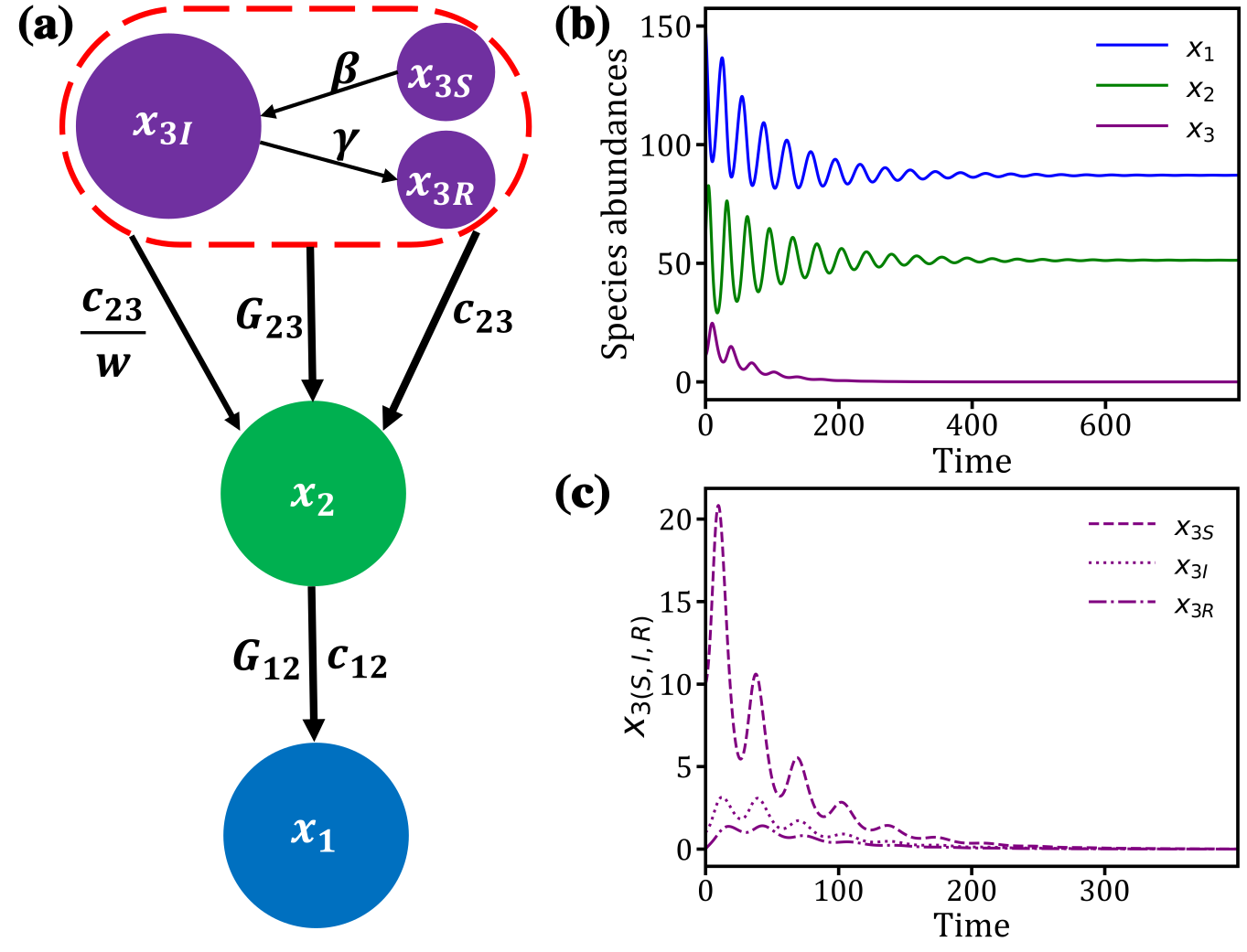}
    \caption{
\textbf{(a)} Schematic representation of the tri-trophic food chain with infection introduced at the tertiary trophic level. The apex predator population is divided into susceptible ($x_{3S}$), infected ($x_{3I}$), and recovered ($x_{3R}$) classes . Infection and recovery occur at rates $\beta$ and $\gamma$, respectively. The interaction strength between the tertiary and secondary trophic levels is modified through the infection-modulated predation parameter $w$.
\textbf{(b)} Time series of the primary species ($x_1$), secondary species ($x_2$), and total tertiary population ($x_3$). Following the introduction of infection at the tertiary level, the apex predator population declines and eventually approaches extinction, while the lower trophic levels converge to stable population values. The values of the infection and recovery rates were taken as $\beta = 0.3$ and $\gamma = 0.1$, with the infection-modulated predation parameter $w = 1.5$.
\textbf{(c)} Time series of the susceptible ($x_{3S}$), infected ($x_{3I}$), and recovered ($x_{3R}$) compartments of the tertiary species. All three epidemiological classes decay over time, indicating the eventual extinction of the tertiary population. The parameter values are listed in Table~\ref{tab:parameters}.
}
\label{fig:Third_species_schematic}
\end{figure}

To investigate the effect of infectious disease on apex predator dynamics, we extend the gravity-based tri-trophic food-chain model by introducing disease transmission within the tertiary trophic level while keeping the primary resource ($x_1$) and the intermediate consumer ($x_2$) disease free. The tertiary population is partitioned into three epidemiological compartments: susceptible ($x_{3S}$), infected ($x_{3I}$), and recovered ($x_{3R}$), as illustrated in Fig.~\ref{fig:Third_species_schematic}(a). The total tertiary population is therefore

\begin{equation}
N_3=x_{3S}+x_{3I}+x_{3R}.
\end{equation}

The ecological consequences of infection are incorporated through the infection-modulated interaction parameter $w$, which characterizes the reduction in predatory performance of infected apex predators. As infection progresses, infected predators become less efficient at capturing and utilizing prey. Accordingly, the effective trophic interaction coefficients are modified as

\begin{equation}
c_{32}^{\mathrm{eff}}(I)=\frac{c_{32}(I)}{w},
\qquad
c_{23}^{\mathrm{eff}}(I)=\frac{c_{23}(I)}{w},
\end{equation}

such that increasing $w$ weakens both prey capture and biomass conversion by infected predators.

Using the gravity-based interaction defined in Eq.~(\ref{eq:gravity_interaction}), the eco-epidemiological dynamics are governed by

\begin{equation}
\begin{aligned}
\dot{x_{1}} &= x_1\left(b_1 - d_1 x_1\right)
- c_{12}\mathcal{I}_{1,2},
\\[1em]
\dot{x_{2}} &= -d_{2}x_{2}
+ c_{21}\mathcal{I}_{2,1}
- c_{23}(S)\mathcal{I}_{2,3S}
- c_{23}^{\mathrm{eff}}(I)\mathcal{I}_{2,3I}
- c_{23}(R)\mathcal{I}_{2,3R},
\\[1em]
\dot{x_{3S}} &= -d_{3S}x_{3S}
+ c_{32}(S)\mathcal{I}_{3S,2}
-\beta\frac{x_{3S}x_{3I}}{N_3}
+c_{32}(R)\mathcal{I}_{3R,2},
\\[1em]
\dot{x_{3I}} &= -d_{3I}x_{3I}
+c_{32}^{\mathrm{eff}}(I)\mathcal{I}_{3I,2}
-\gamma x_{3I}
+\beta\frac{x_{3S}x_{3I}}{N_3},
\\[1em]
\dot{x_{3R}} &= -d_{3R}x_{3R}
+\gamma x_{3I}.
\end{aligned}
\label{eq:third_species_infected}
\end{equation}

Disease transmission follows frequency-dependent incidence. Susceptible apex predators become infected at a rate $\beta x_{3S}x_{3I}/N_3$, while infected individuals recover at rate $\gamma x_{3I}$ and enter the recovered compartment. Mortality of infected predators is incorporated through the parameter $d_{3I}$. Consequently, the epidemiological transitions are fully coupled to the gravity-mediated trophic interactions through the modified interaction coefficients.

\subsubsection{Infection drives apex predator extinction}
We first examine the long-term dynamics following the introduction of infection into the apex predator population. Numerical simulations reveal that infection progressively reduces the abundance of the tertiary predator, ultimately leading to its extinction (Fig.~\ref{fig:Third_species_schematic}(b)). As the predator population collapses, the primary resource and the intermediate consumer converge to non zero fixed population levels. The susceptible, infected, and recovered compartments of the tertiary population all decay to zero (Fig.~\ref{fig:Third_species_schematic}(c)), indicating that disease-induced mortality together with the reduced trophic efficiency of infected predators prevents long-term persistence of the apex predator.

\subsubsection{Gravity-mediated interactions reshape the extinction boundary}
\begin{figure}[htbp]
    \centering
    \includegraphics[width=0.85\linewidth]{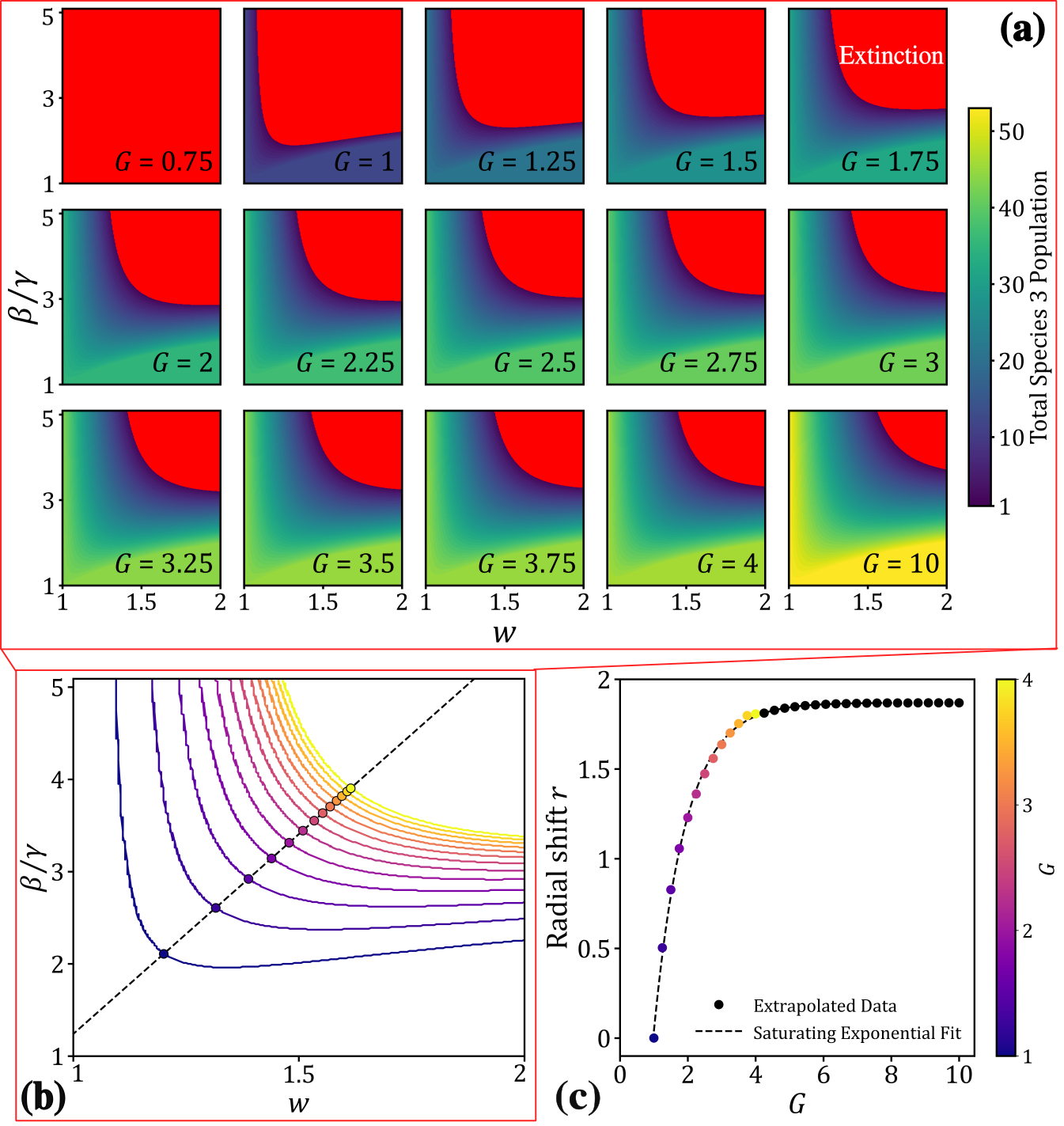}
    \caption{
\textbf{(a)} Parameter space heat maps in the $(w,\beta/\gamma)$ plane showing the long-term population of the apex predator (tertiary trophic level) for different values of the gravity-based coupling strength $G$. Each subplot corresponds to a distinct value of $G$. The color scale represents the population of the tertiary species. The red dotted curve denotes the numerically extracted extinction boundary, separating extinction and persistence regimes. As $G$ increases across the subplots, the region enclosed by the extinction boundary progressively contracts, indicating that stronger gravity-based coupling reduces the parameter combinations leading to tertiary species extinction.
\textbf{(b)} Superposition of the extinction boundaries for all considered values of $G$, together with the curve obtained by joining the knee points of each boundary. The near alignment of these knee points along a direction of approximately $77^\circ$ indicates that the boundaries are effectively tilted and translated versions of one another, revealing a systematic geometric transformation induced by increasing coupling strength.
\textbf{(c)} Radial displacement $r(G)$ of the extinction boundary knee points measured relative to the reference configuration at $G = 1$. Colored markers denote numerically computed intersection points, with color encoding the corresponding value of $G$ as indicated by the color bar. Black markers represent extrapolated values obtained from the fitted model. The solid curve corresponds to a saturating exponential fit, capturing the nonlinear dependence of the radial shift on $G$. The trend demonstrates asymptotic saturation, indicating that the geometric displacement of the extinction boundary approaches a constant magnitude for sufficiently large coupling strength.
}
\label{fig:third_species_Figure_5}
\end{figure}

To determine how gravity-mediated interactions influence predator persistence, we construct extinction diagrams in the $(\beta/\gamma,w)$ parameter space for different values of the gravity coupling strength $G$ (Fig.~\ref{fig:third_species_Figure_5}(a)). The red boundary separates parameter combinations leading to predator persistence from those resulting in extinction. Increasing the gravity coupling systematically contracts the extinction region, indicating that stronger gravity-mediated trophic interactions enhance the resilience of the apex predator against disease-induced extinction. The displacement of the extinction boundary is quantified by the radial shift which exhibits a saturating dependence on the gravity coupling strength (Fig.~\ref{fig:third_species_Figure_5}(c)) (\ref{ Tertiary_model} and \ref{Geometric characterization of the extinction boundary-Third species infected} for detailed explanation). This saturation indicates that the influence of gravity on the extinction threshold becomes progressively weaker as the coupling strength increases.

\subsubsection{Determination of the extinction boundary}
\label{ Third_species_extinction}

To determine the extinction boundary, the eco-epidemiological model is numerically integrated over the $(\beta/\gamma,w)$ parameter space for different values of the gravity coupling strength $G$. For each parameter combination, the asymptotic abundance of the tertiary trophic level is evaluated after eliminating transient dynamics. Parameter combinations yielding a vanishing asymptotic predator population are classified as extinction states, whereas positive asymptotic populations are identified as persistence states.

The extinction boundary separating these two regimes is extracted numerically for each value of $G$. To quantify its geometric evolution, the knee point of each boundary is identified using the maximum-distance method described in ~\ref{Geometric characterization of the extinction boundary-Third species infected}. The displacement of the knee point is characterized by the radial distance

\begin{equation}
r(G)=A\left(1-e^{-G}\right)+C,
\label{eq:radial}
\end{equation}

where $A$ and $C$ are fitting parameters obtained using nonlinear least-squares regression.

\subsubsection{Geometric characterization of the extinction boundary}
\label{Geometric characterization of the extinction boundary-Third species infected}

To quantify this translation, we first determine the knee point of each extinction boundary. For a fixed $G$, the minimum $(w^*(G), \beta^*(G))$ is defined as the point at which the boundary attains its lowest position in parameter space. These minima provide a natural geometric reference for each parabola.

When the minima corresponding to all values of $G$ are connected, they align approximately along a straight line. A linear fit through these minima yields a dominant geometric direction with orientation
\begin{equation}
\theta = 77^\circ.
\end{equation}
This angle represents the effective tilt of the family of extinction parabola. In rotated coordinates defined by
\begin{align}
u &= w\cos\theta + \beta\sin\theta, \\
v &= -w\sin\theta + \beta\cos\theta,
\end{align}
the extinction boundaries appear as approximately parallel shifted curves, confirming that increasing $G$ primarily induces translation rather than deformation.

Using this dominant direction, we construct the straight line passing through the sequence of minima at angle $77^\circ$, as illustrated in Fig.~\ref{fig:third_species_Figure_5}(b). The extinction boundary corresponding to $G=1$ is taken as the reference configuration. The radial displacement of each subsequent parabola is then defined relative to the minimum of the $G=1$ boundary.

For the reference curve corresponding to $G=1$, the minimum point is denoted by
\[
P_0=(w^*(1),\beta^*(1)).
\]
For every other value of $G$, the corresponding minimum point is
\[
P_G=(w^*(G),\beta^*(G)).
\]
The displacement vector relative to the reference configuration is therefore
\[
\Delta P(G)=P_G-P_0.
\]
The magnitude of this displacement defines the radial shift
\[
r(G)=\|\Delta P(G)\|
=
\sqrt{
\left[w^*(G)-w^*(1)\right]^2
+
\left[\beta^*(G)-\beta^*(1)\right]^2
}.
\]

The radial displacement is found to increase with the gravity-based coupling strength and is well described by the saturating exponential function
\begin{equation}
r(G)=A\left(1-e^{-G}\right)+C,
\end{equation}
where $A$ represents the saturation amplitude and $C$ denotes the radial shift at the reference configuration. With fitted parameters
\begin{equation}
r(G) = 5.235489 \left(1 - e^{-G}\right) - 3.351816,
\end{equation}
and coefficient of determination
\begin{equation}
R^2 = 0.997654.
\end{equation}

These results demonstrate that increasing $G$ does not substantially alter the intrinsic curvature of the extinction boundary. Instead, the family of extinction curves behaves as a set of tilted and shifted parabolae that translate coherently along a fixed geometric direction ($\theta = 77^\circ$). The fitted model implies a finite asymptotic displacement,
\begin{equation}
\lim_{G \to \infty} r(G) = A + C,
\end{equation}
showing that the radial shift approaches a constant value for large $G$. Consequently, the contraction of the extinction region saturates, indicating that beyond moderate coupling strengths there is no further significant reduction of the parameter space associated with extinction of the apex predator. 

\begin{figure}[H]
    \centering
    \includegraphics[width=0.9\linewidth]{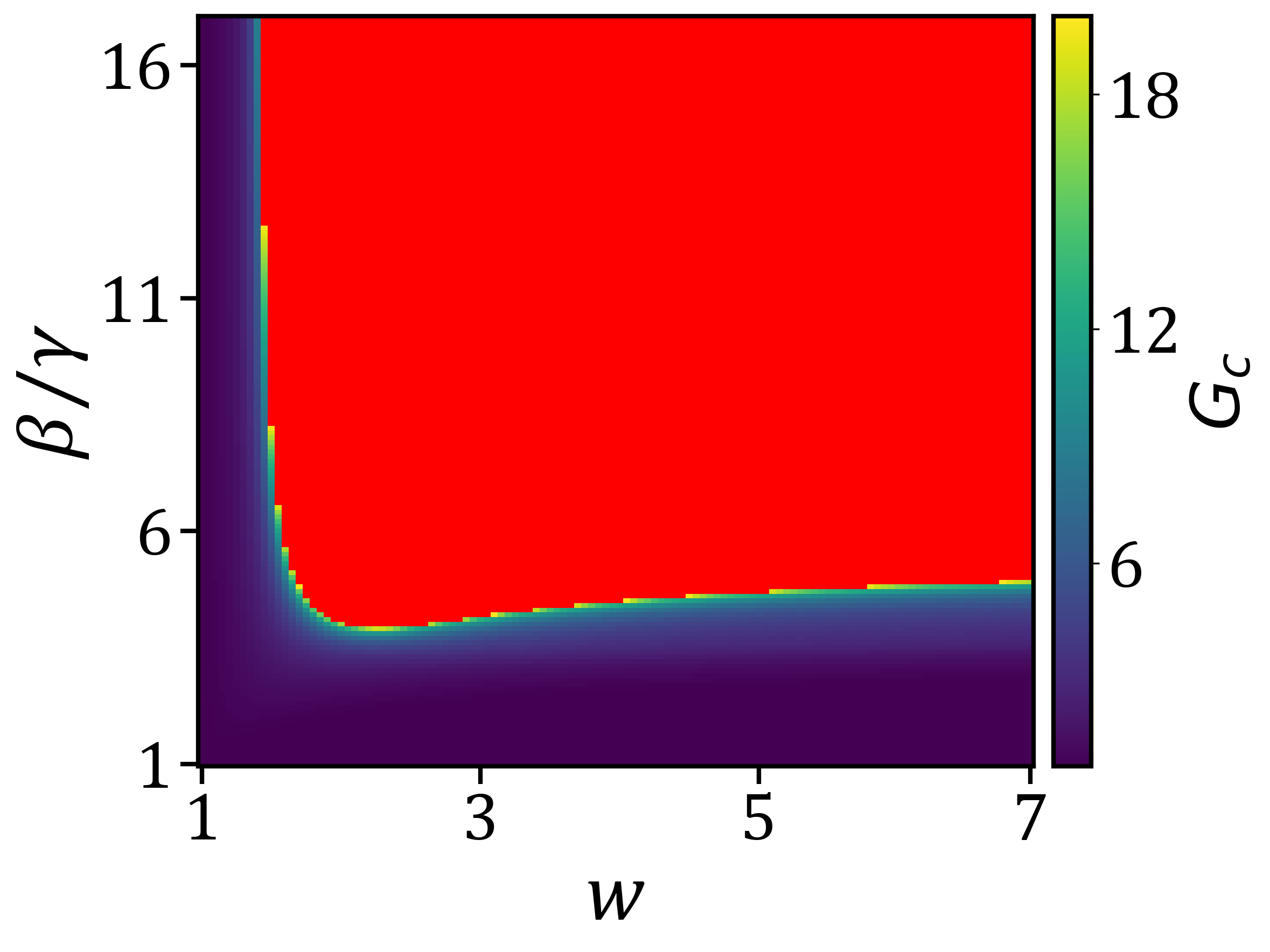}
    \caption{ The color bar represents the value of \(G_{\mathrm{c}}\), defined as the minimum value of $G$ required for the persistence of the tertiary species, (i.e., $x_3 \geq 1$). Red regions indicate parameter regimes for which no finite \(G_{\mathrm{c}}\) exists, for the species to persist. }
    \label{fig:Gcritical_heatmap}
\end{figure}

\subsubsection{Critical gravity coupling determines predator persistence}
To quantify the minimum interaction strength required for $x_3$ to survive, we define the critical gravity coupling strength, $G_{\mathrm c}$, as the threshold separating extinction from persistence of the tertiary species. Here, persistence is defined by an asymptotic abundance satisfying $x_3 \geq 1$. For each combination of  $w$ and $\beta/\gamma$, we evaluate the asymptotic abundance of $x_3$ over the investigated range of $G$ and identify the smallest value of $G$ for which the persistence criterion is satisfied. Figure~\ref{fig:Gcritical_heatmap} shows the resulting distribution of $G_{\mathrm c}$ across the $(w,\beta/\gamma)$ parameter space. The variation in $G_{\mathrm c}$ indicates how the interaction strength required for $x_3$ persistence depends jointly on $w$ and $\beta/\gamma$. Regions with finite $G_{\mathrm c}$ correspond to parameter combinations for which increasing the gravity coupling strength eventually maintains $x_3$ above the persistence threshold. In contrast, the red regions indicate parameter combinations for which no finite $G_{\mathrm c}$ is identified within the investigated range of $G$, such that increasing the coupling strength over this range does not produce persistence of $x_3$. The phase diagram further shows that $G_{\mathrm c}$ increases with both $\beta/\gamma$ and $w$. Thus, parameter combinations associated with stronger epidemiological pressure and greater infection-mediated modification of trophic interactions require progressively stronger gravity-mediated interactions to maintain the tertiary species. 

\begin{figure}
    \centering
    \includegraphics[width=1\linewidth]{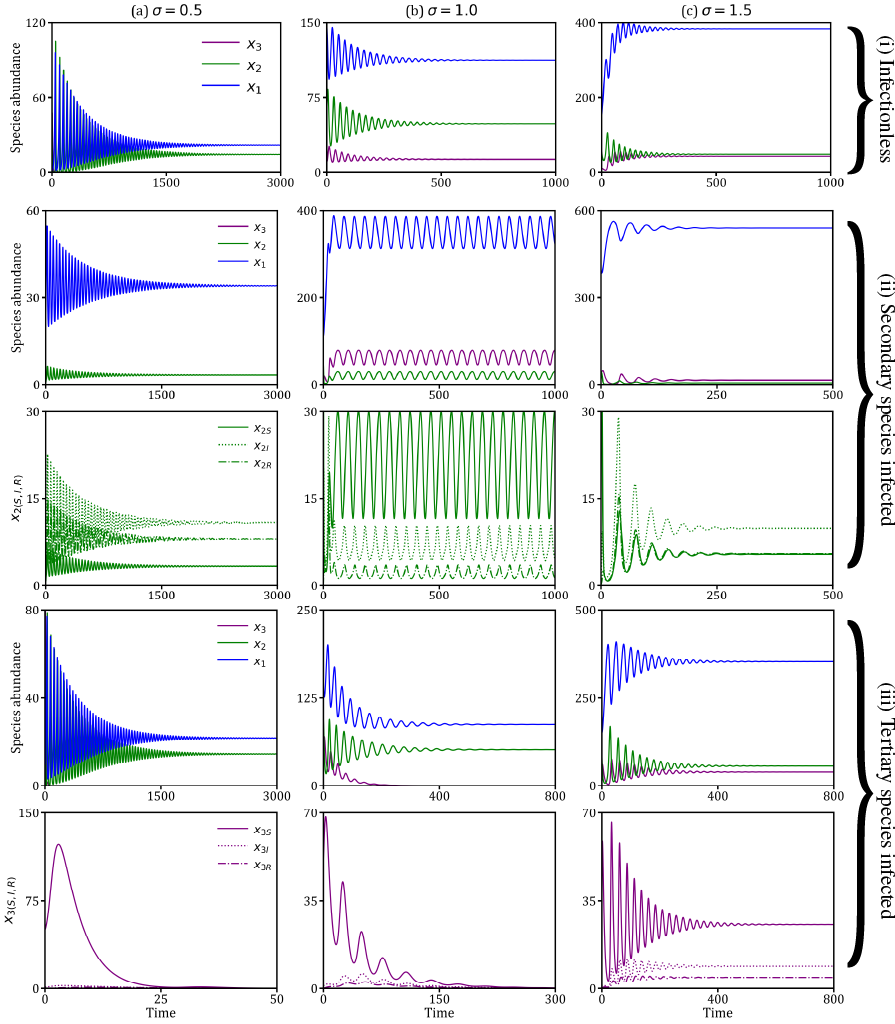}
    \caption{\textbf{Temporal dynamics of species abundances under variation in the interspecific distance ratio $\sigma=a_{12}/a_{23}$.} Columns (a-c) correspond to $\sigma=0.5$, $1.0$, and $1.5$, respectively. Rows (i-iii) represent the infectionless state, secondary species infected, and tertiary species infected, respectively. The solid curves in the species-abundance panels denote the abundances of species 1 ($x_1$), species 2 ($x_2$), and species 3 ($x_3$). The corresponding infection-state panels show the susceptible, infected, and recovered compartments of the infected species, with $x_{2S}$, $x_{2I}$, and $x_{2R}$ for secondary-species infection and $x_{3S}$, $x_{3I}$, and $x_{3R}$ for tertiary-species infection.}
    \label{fig:sigma}
\end{figure}

\section{Effect of the interspecific distance ratio on population and infection dynamics}
\label{supp:sigma_analysis}

To examine the role of spatially mediated trophic interactions, we vary the interspecific distance between the first and second trophic levels while keeping $a_{23}$ fixed. We characterize this effect using the interspecific distance ratio $\sigma=a_{12}/a_{23}$, considering $\sigma=0.5$, $1.0$, and $1.5$. The analysis is performed for the infection-free system, infection at the secondary species, and infection at the tertiary species. The corresponding temporal dynamics are shown in Fig.~\ref{fig:sigma}.

In the infection-free system, increasing $\sigma$ substantially changes both the transient dynamics and the equilibrium abundances of the three trophic levels (Fig.~\ref{fig:sigma}(i)). At $\sigma=0.5$, the populations exhibit pronounced transient oscillations followed by relaxation toward relatively low abundances, with the tertiary species $x_3$ eventually becoming extinct. Increasing the ratio to $\sigma=1.0$ results in faster decay of the transient oscillations and substantially higher equilibrium abundances of $x_1$, $x_2$, and $x_3$. At $\sigma=1.5$, the three populations approach still higher equilibrium abundances, while the transient oscillations are considerably reduced in both amplitude and duration. Thus, increasing the relative distance $a_{12}$ modifies the strength of the interaction between the lower trophic levels and consequently reorganizes both the equilibrium structure and transient relaxation of the infection-free food chain.

The effect of $\sigma$ becomes more pronounced when infection is introduced at the secondary species (Fig.~\ref{fig:sigma}(ii)). At $\sigma=0.5$, $x_1$ and $x_2$ exhibit damped oscillations, while $x_3$ eventually becomes extinct. The susceptible, infected, and recovered compartments of the secondary species, $x_{2S}$, $x_{2I}$, and $x_{2R}$, display corresponding transient oscillations that decay over time. At $\sigma=1.0$, the system exhibits a qualitatively different dynamical regime, with sustained large-amplitude oscillations in $x_1$ accompanied by persistent oscillations in $x_2$ and $x_3$. The infection compartments also oscillate persistently, indicating that the change in interspecific distance can modify the temporal behavior of the epidemic together with the trophic dynamics. At $\sigma=1.5$, $x_1$ rapidly approaches a higher equilibrium abundance, whereas the oscillations in $x_2$ and $x_3$ are comparatively weak and decay toward stationary states. The infection compartments similarly exhibit damped oscillations and approach low steady-state abundances. These results indicate that changes in $a_{12}$, even with $a_{23}$ fixed, can shift the system between qualitatively different transient and persistent dynamical regimes.

When infection occurs at the tertiary species, variation in $\sigma$ again produces substantial changes in both the trophic and infection dynamics (Fig.~\ref{fig:sigma}(iii)). At $\sigma=0.5$, the total populations show pronounced but progressively damped oscillations. The susceptible tertiary population $x_{3S}$ undergoes a strong initial outbreak followed by a decline toward extinction, while the infected and recovered compartments, $x_{3I}$ and $x_{3R}$, also decline relatively rapidly. At $\sigma=1.0$, the primary species reaches a higher equilibrium abundance relative to the secondary species, whereas the tertiary population eventually becomes extinct. The infection dynamics show successive peaks in $x_{3S}$ with decreasing amplitude, followed by a decline in the infected and recovered compartments. At $\sigma=1.5$, the primary species reaches its highest equilibrium abundance among the three distance ratios considered, while $x_2$ and $x_3$ exhibit damped oscillations toward stationary states. The tertiary infection dynamics also differ markedly: $x_{3S}$ exhibits successive oscillatory peaks that gradually decay toward a nonzero level, accompanied by smaller responses in $x_{3I}$ and $x_{3R}$.

Overall, the results demonstrate that the interspecific distance ratio $\sigma$ acts as an important control parameter for the coupled trophic and epidemiological dynamics. Although $a_{23}$ remains fixed, changing $a_{12}$ alters the relative strength of trophic coupling and thereby changes equilibrium abundances, transient relaxation, oscillatory persistence, and tertiary-species persistence. The effect is particularly evident when infection is present, where the same change in $\sigma$ produces distinct outbreak trajectories and can determine whether the tertiary species persists or becomes extinct. These results show that spatially mediated interaction distances cannot be treated independently of the epidemiological dynamics when characterizing the stability and persistence of the tri-trophic system.

\section{Discussion}
Our results demonstrate that the ecological consequences of infectious disease depend strongly on both the trophic position at which disease is introduced and the strength of gravity-mediated trophic interactions. Starting from a stable disease-free tri-trophic food chain, introducing infection into different trophic levels leads to fundamentally different ecological outcomes. Infection of the intermediate consumer destabilizes the coexistence equilibrium through sustained oscillations, whereas infection of the apex predator primarily reduces predator persistence and can ultimately lead to extinction. These contrasting responses show that the same epidemiological process can produce distinct ecosystem-level transitions depending on where disease enters the trophic hierarchy.

The different responses arise from the contrasting ecological roles of the infected species. The intermediate consumer links the primary resource to the top predator and therefore mediates biomass transfer across the entire food chain. Infection at this level simultaneously alters resource consumption and predator support, creating coupled ecological and epidemiological feedback that destabilize the coexistence equilibrium through a Hopf bifurcation. In contrast, infection of the top predator primarily weakens predation and biomass acquisition at the terminal trophic level. Consequently, the dominant ecological response shifts from oscillatory instability to predator extinction. These findings suggest that trophic position is a fundamental determinant of how disease propagates through ecological communities.

An important outcome of this work is the identification of the gravity coupling strength as a regulator of disease-driven ecological transitions. By systematically varying the gravity coupling, we show that the thresholds separating stable coexistence, oscillatory dynamics, and predator extinction shift in a predictable manner. Increasing the gravity coupling promotes oscillatory behavior when infection occurs at the intermediate trophic level, whereas it enhances predator persistence when infection affects the top trophic level. Thus, the gravity coupling does not alter the nature of the ecological transitions themselves, but instead regulates the conditions under which they occur.

Beyond identifying these transitions, our analyses provide quantitative measures for characterizing ecosystem resilience. The geometric evolution of the oscillatory and extinction boundaries demonstrates how the stability landscape reorganizes as the gravity coupling changes, while the critical gravity coupling strength provides the minimum interaction strength required for long-term persistence of the apex predator. Together, these quantities complement conventional bifurcation analysis by describing how ecological transitions evolve across parameter space rather than only identifying their onset.

Although the present study considers a minimal deterministic tri-trophic food chain, the proposed gravity-based framework is sufficiently general to be extended to more realistic ecological systems. Future work could incorporate spatial heterogeneity, diffusion based movement of species between two-habitats, environmental stochasticity, adaptive behavior, multiple pathogens, or complex food-web architectures to investigate how gravity-mediated trophic interactions influence ecosystem stability under more realistic ecological conditions. Such extensions would facilitate direct comparisons with empirical systems and help assess the broader applicability of the framework.

Overall, this study establishes a gravity-based eco-epidemiological framework for investigating how infectious disease interacts with trophic organization. Our results show that the trophic location of disease determines the type of ecological transition, whereas the gravity coupling strength regulates the resilience of these transitions. We expect that this framework will provide a useful foundation for studying disease-driven ecological dynamics in food chains and more complex ecological communities.

\clearpage
\bibliography{ref}
\bibliographystyle{unsrt}
\end{document}